\documentclass[sigconf]{aamas}

\usepackage{graphicx}
\usepackage{algorithm}
\usepackage{algorithmic}
\usepackage{subcaption}
\usepackage{makecell}
\usepackage{tabularx}

\usepackage{balance} 

\acmSubmissionID{869}

\title[AAMAS-2025 Formatting Instructions]{On the Parallels Between Evolutionary Theory and the State of AI}

\author{Zeki Doruk Erden}
\affiliation{
  \institution{École Polytechnique Fédérale de Lausanne}
  \city{Lausanne}
  \country{Switzerland}}
\email{zeki.erden@epfl.ch}

\author{Boi Faltings}
\affiliation{
  \institution{École Polytechnique Fédérale de Lausanne}
  \city{Lausanne}
  \country{Switzerland}}
\email{boi.faltings@epfl.ch}

\begin{abstract}
This article critically examines the foundational principles of contemporary AI methods, exploring the limitations that hinder its potential. We draw parallels between the modern AI landscape and the 20\textsuperscript{th}-century Modern Synthesis in evolutionary biology, and highlight how advancements in evolutionary theory that augmented the Modern Synthesis, particularly those of Evolutionary Developmental Biology, offer insights that can inform a new design paradigm for AI. By synthesizing findings across AI and evolutionary theory, we propose a pathway to overcome existing limitations, enabling AI to achieve its aspirational goals.
\end{abstract}


\newcommand{\BibTeX}{\rm B\kern-.05em{\sc i\kern-.025em b}\kern-.08em\TeX}

\begin{document}


\pagestyle{fancy}
\fancyhead{}


\maketitle

\section{Introduction}
\label{sec:introduction}

Over the past decade, artificial intelligence has increasingly become part of daily life, significantly affecting the average person \cite{precedenceresearchArtificialIntelligence, mcelheran2024ai, springsappsLargeLanguage}. Can we confidently say that the current AI design paradigm—overparameterized neural networks optimized through gradient descent—will lead to reliable, continuously improving AI systems? Are existing technologies truly meeting expectations, or do challenges like continual learning, opacity, and symbolic reasoning reveal limitations that necessitate a paradigm shift? How do these issues connect to advancements in evolutionary theory?

This article addresses these questions by first overviewing contemporary AI technologies, analyzing the strengths and weaknesses of the last decade's machine learning paradigm (Section \ref{sec:contemporary_ai}). Next, we will draw parallels between the dominant view of evolutionary biology from the 20\textsuperscript{th} and contemporary machine learning—specifically, between explanatory limitations of the former and capability shortcomings of the latter. We will explore significant advancements in evolutionary theory that relate to today’s AI landscape and conclude by outlining what the future of AI should encompass to overcome current limitations and realize the technology's full potential  (Sections \ref{sec:evolution} and \ref{sec:ai_future}).

This paper is structured as a theoretical and conceptual argumentative work, integrating a targeted review of relevant advances in evolutionary theory as they relate to our arguments, an interpretation of these advances through the lens of contemporary machine learning technologies, and a proposed pathway for the future of AI from the developed perspective. While explicit, exemplifying methods based on this theory are not included in the main text—both to maintain our focus on the argumentative side and to avoid constraining it to specific implementations—references to and discussions of existing works that apply these principles are provided in Section \ref{sec:ai_future} for readers seeking concrete examples.

\section{Contemporary Machine Learning}
\label{sec:contemporary_ai}

Methods from the previous AI design paradigm, known as "classical" or "symbolic AI," often struggled with complex, high-dimensional tasks. However, as a consequence of advancements in computing hardware \cite{khan2021advancements, baji2017gpu} and the availability of domain-specific data \cite{duan2019artificial} making training large-scale neural networks practical, many tasks once deemed unsolvable with classical methods can now be effectively learned using neural networks given sufficient training data \cite{khan2021machine, chai2021deep,zhao2023survey, min2023recent, li2017deep}. This is largely due to neural networks' ability to approximate any function if they have a sufficiently large number of parameters \cite{cybenko1989approximation}. In practice, this often means heavily overparameterized networks \cite{du2018power}, resulting in many more parameters than theoretically necessary, as gradient descent requires ample degrees of freedom to navigate the solution landscape effectively.

As AI steadily conquers challenges once thought insurmountable, growing attention is being drawn to fundamental limitations in the principles underlying current algorithms and their ability to integrate into broader systems \cite{clune2019ai,zador2019critique,marcus2018deep,lecun2022path}. We think that the major limitations can be categorized in two broad categories.

\subsection{Obliteration of the existing knowledge when learning new information}
\label{sec:ai_destructive_adaptation}

Modern ML methods are incapable of effectively integrating new knowledge into existing knowledge without destroying old information, which is crucial for continual learning \cite{vandeven2024continuallearningcatastrophicforgetting, hadsell2020embracing}. This limitation not only forces AI systems to undergo extensive retraining and reinteraction with the environment where such interaction is not feasible, but also eliminates their prospects to expand their knowledge and capabilities in an adaptive and recursive manner.

How old knowledge is destroyed when neural networks learn new tasks will be crucial for our argument: Neural networks (NNs) approximate a function that represents the average effect of a large dataset, which is only aligned with the desired essence of the task when all relevant data is considered together. In large, overparameterized networks, this enables impressive performance, provided the data is available from the outset. However, when a new task is introduced, the network does not create a combined objective function; instead, it optimizes exclusively the new target function, with no pressure to retain performance on previous tasks. This leads to the destruction of previously learned knowledge (represented by existing weight patterns) when faced with new tasks, manifesting as a sharp decline in performance on the old task \cite{vandeven2024continuallearningcatastrophicforgetting}.\footnote{This phenomenon is often called "catastrophic forgetting" in AI literature, but this term is misleading; it suggests a gradual loss of unused knowledge, while the actual process is the active destruction of existing knowledge when learning a new task. To avoid this, we refer to this phenomenon as "destructive adaptation" instead.}

Several \textit{ad-hoc} fixes exist for this problem of \textit{destructive adaptation}. The most straightforward involves storing and replaying past examples \cite{rolnick2019experience, buzzega2021rethinking, buzzega2020dark, galashov2023continually}. While simple, this approach merely circumvents the issue: As a system continues to learn, it must retain more past data, leading to explosive memory, computational demands and dilution of previous knowledge. Such methods may work in controlled experiments, but they are impractical for real-world agents which can be expected to face millions of distinct situations over their lifetime. Other approaches try to mitigate the issue at a higher level but depend on restrictive assumptions, like clear task boundaries \cite{masse2018alleviating, jacobson2022task, kirkpatrick2017overcoming, wang2022continual, iyer2022avoiding}, external signals indicating the active task \cite{rusu2016progressive}, access to bulk of relevant data at once \cite{erden2024directed, lee2020neural, iyer2022avoiding} or applying only to tasks with high orthogonality among them \cite{iyer2022avoiding,damleactive}. These assumptions clash with the structure of real-world learning, where agents encounter information sequentially, integrating new knowledge with past experiences without the need to store all prior data or rely on neatly defined 'tasks.'

\subsection{Incomprehensibility \& non-engineerability}
\label{sec:ai_incomprehensible}

The second critical limitation of modern ML methods is the incomprehensibility of AI systems' internal structures, which hinders their widespread adoption and poses significant risks \cite{marcus2018deep}. NNs rely on overparameterization, deriving representations from continuous parameters fine-tuned to a few significant digits and exhibiting nonlinear responses from hundreds or thousands of inputs per neuron. This yields a learned structure that is \textit{mostly} incomprehensible—and the narrow window implied by this 'mostly' is typically limited to very low-dimensional problems or demands tedious analysis, often yielding only vague and relatively low-resolution insights into what has actually been learned \cite{naseer2021intriguing}. Currently, no mainstream method reliably incorporates a modular and hierarchical structure to enhance comprehensibility, with existing attempts \cite{su2024focuslearn, goyal2020object, pateria2021hierarchical} still not addressing the internal complexity of overparameterized, continuous, and nonlinear components. Research on explaining internal operations mainly focuses on \textit{post-hoc} analyses of responses and parameter statistics rather than addressing the systems' intrinsic complexity \cite{xu2019explainable} (hence, such methods are often referred to as 'explainable AI' rather than 'comprehensible AI'). This sharply contrasts with nearly all human-engineered systems. While it’s unrealistic to expect a system trained on vast data to represent all that knowledge with a small number of parameters or to be immediately comprehensible at every level, it is reasonable to desire some level of comprehensibility at higher levels of abstraction or within each organizational layer. This lack of clarity hampers our understanding of how these systems function and inhibits decomposability and low-level modifiability, making AI inherently uncontrollable and unreliable beyond its statistical guarantees.

A related limitation of current machine learning systems is their hindered integration with human-designed algorithms and established knowledge frameworks, due to the aforementioned inherent non-decomposability and lack of structured organization in the learned models. This includes explicit information about the environment, such as pre-existing maps, and classical AI methods like search, logic, planning, and constraint satisfaction \cite{norvig2002modern}.\footnote{In saying this, we primarily refer to the inability of an agent, embodied in a physical or virtual environment, to perform the relevant processes or utilize a learned environment model \textit{directly}. In doing so, we distinguish this capability from recent works that treat planning as a language-processing problem, which we regard as a different approach that operates in a distinct domain of application \cite{ruan2023tptu, shi2025tool}.} Integrating these methods is desirable, as classical AI has decades of research backing it and offers robust techniques for tasks that data alone cannot efficiently solve. These methods provide benefits like precision, reliability, and comprehensibility, which the current machine learning paradigm cannot leverage due to its inherent limitations.

An example of this challenge is in deliberative behavior and planning, which enable an AI agent to make goal-driven decisions and avoid undesired outcomes without needing to re-interact with the world to learn desirable behavior, and backed by extensive planning literature \cite{ghallab2016automated}. Despite considerable research, no mainstream method effectively integrates deliberative processes with learning systems in a targeted, goal-oriented manner (as opposed to those based on forward sampling \cite{mcmahon2022survey, otte2015survey}) akin to traditional planning, largely due to the monolithic, black-box nature of neural networks' environment models. Some approaches attempt to circumvent this limitation by embedding goal-directed behavior into the learning process, conditioning on various goals during training \cite{colas2019curious, colas2022autotelic, liu2022goal}. This is undoubtedly valuable in the context of contemporary machine learning. However, it sacrifices one of the key motivations of deliberative behavior in ‘symbolic’ approaches—namely, the ability to devise intelligent behavior for an entirely new goal within a known environmental model. On the flip side, planning research that attempts to learn environment models often operates under restrictive assumptions \cite{jimenez2012review, mordoch2023learning, verma2021asking, stern2017efficient} that fail to capture the expressiveness of modern learning systems. Integrating expressive environmental model learning with effective planning capabilities would significantly enhance the functionality of agential AI systems.\footnote{A noteworthy approach for integrating capabilities of modern ML systems with the strengths of classical methods is neuro-symbolic AI \cite{wan2024towards, colelough2025neuro}. However, the field is limited as the NNs used are often still black boxes, hindering the benefits of symbolic reasoning, like comprehensibility and precision, from being applied to internal representations, with advantages appearing only at higher design levels. As will be discussed, there is no inherent reason for this limitation with a proper design paradigm.}

\section{Augmented Synthesis of Evolution}
\label{sec:evolution}

Evolutionary theory is usually regarded as a field only tangentially related to AI. However, upon closer examination, it becomes evident that it is arguably the most relevant of all, as this field has traversed a remarkably similar conceptual and intellectual trajectory to that which AI is currently navigating, albeit in previous decades. The dominant view of evolution in the 20\textsuperscript{th} century, Modern Synthesis (MS), has striking parallels to contemporary ML approaches in terms of its understanding of adaptive tasks, the processes of adaptation, and, importantly, its explanatory limitations. Consequently, insights from the recent augmentations to MS in evolutionary theory provide the necessary conceptual framework for overcoming the limitations of today's ML systems. Recognizing that many of our readers may not be well-acquainted with these developments, we offer a detailed overview, followed by an exploration of their relevance to contemporary AI.

\subsection{Modern Synthesis and its limits}
\label{sec:modern_synthesis}

Darwinian theories of selection \cite{darwin1964origin}, combined with Mendelian genetics and modern understanding of genetic variation has formed the understanding of evolutionary biology that has dominated the Modern Synthesis \cite{huxley1942evolution, gilbert2000developmental, mayer2013evolution}. MS views evolution as driven by inheritable \textit{genetic variation}, arising from mutations and recombination, enabling populations to adapt to environmental changes. Selection, acting on \textit{phenotypic variation}, favors traits that enhance survival and reproduction. MS treats evolution as a process at the population level, abstracting away the complex internal processes of organisms and development, which aren't central to its explanation \cite{walsh2015organisms, gilbert2000developmental, marc2005plausibility, carroll2005endless, laland2015extended}. Biological evolution is traditionally understood as a property of populations rather than individual organisms; under the strict Modern Synthesis framework, it is expected to proceed through slow, gradual shifts in a population’s genetic makeup \cite{johnston2019population, carroll2005endless}, with also no inherent reason to anticipate an accelerating rate of phenotypic change over time.

Modern Synthesis has been immensely successful in advancing our understanding of evolution, with many practical implications \cite{gould2009antibiotic, dolgova2018medicine, thrall2011evolution, olivieri2016evolution}. But as powerful as it is, it leaves gaps in its ability to explain aspects of evolution, whose importance are now understood to be much more important than initially thought: Firstly, the MS does not incorporate (neither as explanandums nor as explanans) the generation of biological structures \cite{walsh2015organisms, gilbert2000developmental, marc2005plausibility, carroll2005endless} that exhibit widespread features like repetition/reuse \cite{preston2011reduce, anderson2010neural, barthelemy1991levels}, correlation \cite{paaby2013many, watson2014evolution, price1992evolution}, modularity \cite{lorenz2011emergence, clarke1992modularity, kadelka2023modularity, clune2013evolutionary, wagner2007road}, and hierarchical composition \cite{mengistu2016evolutionary, ingber2003tensegrity, uversky2021networks, grene1987hierarchies, pan2014exploring} in phenotypes. By viewing evolution as a gradual optimization of allele frequencies, MS excludes the internal complexity of organisms from its inquiry, deeming it supposedly irrelevant for understanding evolutionary processes and abstracting it away within genetic variation. This perspective treats organisms more as mere vessels for genes than as intricate entities with their own developmental processes \cite{walsh2015organisms}. Secondly, the qualitative pattern of evolutionary progression predicted by Modern Synthesis differs from what is observed. Evolution is not merely a steady, linear process; it also includes periods of rapid transformation \cite{gould1977punctuated, marc2005plausibility} and more notably, an exponential increase in the speed of morphological change and phenotypic complexity throughout evolutionary history \cite{smith1997major, koonin2007biological, russell1983exponential, sharov2006genome, heylighen2000evolutionary, kurzweil2006singularity}. To exemplify, the timeline of life's complexity—from the emergence of life ($3.5$ bya) to eukaryotes ($2$ bya), multicellular organisms ($1$ bya), and the Homo genus ($3$ mya), see also Table 1 in \cite{gerhart2007theory}—reveals that while these exemplary events occur over a roughly linear timescale, phenotypic complexity has increased exponentially \cite{vosseberg2024emerging, grosberg2007evolution, ispolatov2012division}.

The underlying issue is that the Modern Synthesis effectively addresses two of the foundational pillars of evolutionary theory —selection and inheritance— but provides a limited account of the third pillar, variation \cite{marc2005plausibility}. While it clarifies the origins of \textit{genetic variation}, it treats the transition to phenotypes as a black box, failing to fully explain \textit{phenotypic variation}. This leaves much of the "how" of evolution unaddressed \cite{marc2005plausibility, laland2015extended}.\footnote{An analogy can be drawn between Modern Synthesis and Genetic Algorithms \cite{sivanandam2008genetic}, which our readers may be more familiar with compared to evolutionary biology. These algorithms excel at optimizing predefined structures, such as selecting the best variable combinations or tuning design parameters \cite{bayley2008design, chiroma2017neural}, but they cannot spontaneously generate new structures or implement rapid, large-scale adaptive changes, except a small class of methods called indirect encodings \cite{meli2021study, gauci2010indirect}. See also the related footnote in Section \ref{sec:ai_future_principles}.} In recent decades, advancements in biology have expanded MS, addressing gaps in its statistical optimization framework \cite{laland2015extended}. Among these, evolutionary developmental biology (EDB) has been particularly impactful \cite{carroll2005endless, marc2005plausibility, west2003developmental, gerhart2007theory}, focusing on how developmental mechanisms drive evolutionary patterns. EDB dispels the notion that genotype-to-phenotype transformation is irrelevant to understanding evolution, revealing that universally shared developmental principles not only drive organismal complexity and adaptability but also illuminate key aspects of evolution itself.

\subsection{\textbf{The structure of gene regulation}}
\label{sec:edb_structure_genereg}

A central insight of EDB is that the generation of phenotype is driven not only by the coding genes and proteins an organism possesses, but is primarily shaped by changes in the genes that govern their \textit{regulation} (“where,” “when,” and “how much” a gene is expressed) \cite{carroll2005endless}. Moreover, these regulatory genes can control the expression of other regulatory genes, enabling the hierarchical regulation of entire pathways of gene expression, where master regulators can initiate downstream cascades \cite{hansen2015effects}. This web of interactions, where different parts of the genome activate or suppress one another and coordinate downstream processes, is known as a \textit{gene regulatory network (GRN)} \cite{davidson2010regulatory,levine2005gene}. Small changes in regulation, when unfolded across time during development, can lead to dramatic alterations in the adult organism’s phenotype. A striking example of this is a famous experiment in fruit flies, where a change in pathways that regulate eye development resulted in the development of relatively complete eye-like structures on the wings \cite{halder1995induction}. This serves to emphasize that there is no need to "evolve a new eye from scratch", and that regulatory changes could re-deploy an existing developmental program to form a structure that could be integrated within the new location, allowing for large qualitative changes in phenotype with little modification at the genome \cite{carroll2005endless}.

As highlighted earlier, biological organisms consistently showcase fundamental structural properties like modularity, hierarchy, repetition \cite{preston2011reduce, barthelemy1991levels, lorenz2011emergence, kadelka2023modularity, clune2013evolutionary, ingber2003tensegrity, uversky2021networks, pan2014exploring} and correlated changes \cite{paaby2013many, watson2014evolution, price1992evolution}. These traits stem from the architecture of the regulatory networks previously discussed \cite{carroll2005endless, alcala2021modularity, verd2019modularity, wagner2011pleiotropic}, such as reuse of regulatory mechanisms across various processes (repetition/correlation), distinct gene sets regulating different parts of the organism (modularity), or higher-level regulatory genes controlling the expression of downstream genes (hierarchy). These phenotypic structural properties of organisms reflect the nature and organization of gene expression within the genome. 

Additionally, these structural properties significantly influence how evolution alters organisms, facilitating the development of new functional structures while reducing the likelihood of detrimental changes: Modularity and hierarchical organization facilitate both correlated changes—where modifications to one part of the organism trigger rapid, synchronous changes in others—and isolated changes, allowing alterations in one component without affecting others \cite{carroll2005endless, verd2019modularity, clune2013evolutionary, mengistu2016evolutionary, seki2012evolutionary, eliason2023early}. Repetition of substructures, on the other hand, allows for the rapid generation of new structures from a moderately developed origin \cite{wagner2011pleiotropic, carroll1995homeotic}. Furthermore, they facilitate the preservation of ancestral traits \cite{szilagyi2020phenotypes, carroll2005endless}, by rapid deactivation switch points responsible for expressing these traits without completely dismantling the underlying genetic machinery, potentially allowing for rapid re-emergence of a past, functional trait.

\subsection{\textbf{Process-encoding genome}}
\label{sec:edb_process_varsel}

Some structures or behavioral responses of an organism do not have predictable, pre-determined shapes or patterns and must be generated in response to the demands of environment or the rest of the organism. The important point in understanding how these types of adaptively generated structures can be formed, evolved, and in turn affect the further evolution of the organism lies in the understanding that the genome does not act as a bluprint or an encoding of the organism \cite{natureItsTime}. Instead, it encodes the \textit{cellular processes} that, throughout developmental unfolding, modify the structure, responses, and behaviors of the cells that express them—more like a control system for the lowest-level unit of organization \cite{marc2005plausibility, gerhart2007theory, west2003developmental}.

This structure-generation scheme ensures reliable development under changing conditions, promoting evolutionary adaptability. For instance, in tetrapod limb development, only bone and dermis precursors exist initially, while other tissues form adaptively around them in response to their signals \cite{gerhart2007theory, marc2005plausibility, kardon2003tcf4, christ2002limb, li2012regulation, adair2010angiogenesis, ferrara2003biology}. Consequently, even major skeletal changes, such as the emergence of new appendages \cite{malik2014polydactyly}, do not require simultaneous evolution of associated systems, enabling rapid and stable adaptations \cite{marc2005plausibility, gerhart2007theory, west2003developmental}.

The operation of immune system, based on selective proliferation of lymphocytes that recognize antigens with introduced mutations to enhance specifity, offers an even more striking example of adaptive somatic response, combatting pathogenic threats without evolutionary preprogramming \cite{burnet1957modification, burnet1957modification, rajewsky1996clonal}. This exemplifies a general class of mechanisms often referred to as \textit{exploratory processes} \cite{gerhart2007theory, marc2005plausibility}, which leverage the \textit{variation and selection} principles underlying biological evolution locally, where variants of a substructure are generated and then refined through a selective signal. Similar principles govern neural development \cite{hiesinger2021self, marc2005plausibility}, where synapses \& axons are initially overproduced and later refined through pruning of weaker ones \cite{bourgeois1989synaptogenesis, rakic1986concurrent, chechik1998synaptic, sakai2020synaptic, petanjek2023dendritic, lamantia1990axon, rakic1983overproduction, provis1985human, kaiser2009simple, innocenti1997exuberant}. It is noteworthy that these evolutionary mechanisms underpin intelligence itself — a point we elaborate in Appendix.

We note a perspective linking mainstream machine learning techniques to internal evolutionary processes: Overparameterized neural networks trained through gradient descent can be seen as undergoing a selection process from an initially abundant pool of variation in form of randomly initialized weight patterns. Gradient descent acts as a selective force, amplifying "beneficial" weights that reduce error while diminishing others. This process converges to a minimum where the "fit" configurations stabilize. Importantly, \textit{this variation-selection cycle is not iterative}; there is no mechanism for regenerating variation atop existing structures. Instead, it is a directed selection mechanism transforming an initial variation pool into a stabilized network with non-random, functional weight patterns. New variation cannot be introduced over this adapted structure without disrupting existing knowledge, as any new selective pressure (gradient in a new direction) affects the entire network. This perspective not only provides a new interpretation for the expressive power of NNs—an immense initial reservoir of variation refined by a strong selective signal—but also sheds a new light on the destructive adaptation problem in continual learning: Fundamentally, the issue stems from the network’s inability to generate new variation locally when and where needed, without annihilating prior weight patterns, standing in contrast to exploratory processes in biological systems that can locally generate variation on demand.

\subsection{Implications of EDB for understanding the pattern of evolutionary progression}
\label{sec:edb_understanding}

Findings from EDB suggests that large portions of the expressed genome—specifically coding DNA, which encodes proteins, as well as the associated low-level regulatory networks—are highly conserved across distantly related organisms \cite{marc2005plausibility, gerhart2007theory}. This includes the fundamental processes responsible for generating basic cellular structures \cite{holstein2012evolution}, as well as higher-order structures \cite{lemons2006genomic} and processes that allow organisms to generate adaptive responses \cite{holstein2012evolution}. Such \textit{conserved core processes} \cite{marc2005plausibility} evolved early in evolutionary history and have been preserved across phylogenetic distances. By contrast, major evolutionary changes since these early developments have primarily occurred at the level of regulatory networks, which control the expression and interaction of this foundational core processes. Examples include changes in regulatory elements that drive morphological novelties \cite{burgess2016sonic, tickle2017sonic, rebeiz2017enhancer} and the reconfiguration of neural gene expression networks that enhanced brain function in humans \cite{wang2016divergence, patoori2022young, suresh2023comparative}. These regulatory modifications not only drive organismal diversity but can themselves become conserved and serve as substrate for further evolutionary change. Crucially, the core processes (whether exploratory or downstream structure-generating) are typically connected via a \textit{weak linkage}, where regulatory signals (inputs/conditions) for their activation are very simple, as opposed to complex and precise ones \cite{gerhart2007theory, marc2005plausibility}, greatly facilitating regulatory changes or process recombinations.

Recall our earlier discussion on how the statistical-optimization framework of the Modern Synthesis, taken alone, does not fully capture the complexities of evolutionary history, including phenomena rapid \& accelerating increases in phenotypic complexity and the exponential growth of organismal capabilities. In contrast, an evolutionary theory that incorporates developmental processes offers a better explanation: Once core processes evolve, they can be reused with minimal genetic changes focused on regulating these processes \cite{marc2005plausibility, gerhart2007theory}. This ability to re-deploy core processes, supported by adaptable peripheral systems, leads to significant phenotypic differences with few genetic modifications. Furthermore, a pre-existing repertoire of such core processes can be combined or reconfigured to create more complex functions, as seen in the evolution of the vertebrate eye \cite{lamb2007evolution, lamb2008origin}, which repurposed existing components like photoreceptor cells, dermis, and ancestral neural circuitry. Such evolution requires fewer changes than building structures from scratch. Additionally, adaptive exploratory processes enable organisms to develop functionally despite large-scale changes, allowing for exponential enhancement of their capabilities without needing to wait for peripheral evolutionary changes to support them \cite{west2003developmental}. This view explains why periods of rapid increases in phenotypic complexity are a persistent phenomenon throughout evolutionary history, and why these increases seem to accelerate exponentially when viewed across history of life, providing a more comprehensive view than one based solely on the Modern Synthesis.

\subsection{Modern Synthesis and contemporary AI}
\label{sec:ms_ai}

The Modern Synthesis of evolution and contemporary machine learning exhibit striking parallels, with a compelling analogy to be drawn between the \textit{explanatory assumptions and unresolved gaps} of the former and the \textit{capability constraints and functional shortcomings} of the latter. Both view their respective processes of adaptation fundamentally as statistical optimization processes. They emphasize a numerically defined, averaged final performance (reproductive fitness and reward/error) rather than focusing on the nature of their subjects, treating them as largely unstructured, encapsulating relevant properties in abstract continuous variables (gene frequencies and edge weights) without addressing the deeper meaning behind. While each has been successful in its own right, both are considered incomplete for similar reasons: Modern Synthesis does not fully explain certain aspects of evolutionary progress, namely the speed of adaptation, the inherent adaptability of organisms, or the widely-observed structural properties of biological systems. Analogously, contemporary machine learning faces criticism for its data inefficiency ($\sim$speed of adaptation), inability to leverage past knowledge when learning new tasks without sacrificing previously acquired information ($\sim$inherent adaptability), and opaque internal structure ($\sim$structural properties).

The resemblance between the previously dominant view of evolution and contemporary ML, particularly regarding their limitations, raises an intriguing question: Can the principles behind the recent shift in evolutionary theory—largely driven by EDB, with core ideas explored—also form the foundation for improving AI?\footnote{A further complementary angle to consider in the relationship between evolution and intelligence is the possibility of (natural) intelligence being a process governed directly by evolutionary mechanisms. We detail this perspective in the Appendix.} In the next section, we will discuss how these principles can be translated into design principles for a new paradigm in AI system development.

\section{Insights for the future of AI}
\label{sec:ai_future}

\subsection{Learning structured representations without destructive adaptation}
\label{sec:ai_future_principles}

The "design principles" emerging from EDB, which have propelled evolutionary understanding beyond pure Modern Synthesis, offer a path to overcoming AI’s parallel limitations. These principles lay the foundation for a new learning paradigm—one that can produce comprehensible models within a structured representation and avoid the necessity of destroying existing knowledge when acquiring new information. To outline this path, we identify the following principles as particularly significant and integrable into AI (categorization being interdependent and mutually non-excluding):

\textbf{1) Encapsulation and Core Processes to Realize Modularity and Decoupling:} To achieve effective internal decoupling in AI architectures, learning must prioritize the acquisition and encapsulation of fundamental processes—such as perceptual representations and behavioral patterns—mirroring how biological systems encapsulate core processes to enhance reuse and minimize interference. These processes can serve as building blocks for more complex functions\footnote{Note that this point also ties into the mechanisms underlying major evolutionary transitions \cite{smith1997major} and the emergence of new levels of selection—an aspect we have not explored here but is still foundational for the recent advances in evolutionary theory.} and be interconnected through weak linkages \cite{gerhart2007theory}, allowing for the preservation of internal structure during changes to other processes or higher-level structures. In this framework, algorithms inspired by evolutionary patterns observed in biological systems—e.g. duplication and differentiation processes \cite{taylor2004duplication, copley2020evolution, minelli2000limbs} or local variation-selection—can conveniently modify components within previously learned structures, or generate variants tailored for specialized purposes, while preserving the integrity of the rest. 

\textit{\textbf{Contrast:}} This contrasts with the complete model or behavior being learned as a flat, non-decomposable system in modern NNs, as well as the inherent assumption of having access to the entire dataset related to the full task during learning (see Section \ref{sec:ai_destructive_adaptation}).

\textbf{2) Higher-Level "Regulatory" Processes for Selective Reuse and Hierarchical Organization:} Similar to how regulatory mechanisms in biology govern the activation of developmental pathways by selectively deploying core processes (Sec. \ref{sec:edb_structure_genereg}), AI learning systems should integrate mechanisms that regulate the higher-level deployment of their processes (1). Learning should potentially be able to procees via adjustments to the "regulatory" high-level processes that oversee the activation of potentially complex downstream core processes, without necessitating modifications to the internal operations of these core processes, if the task can be accomplished through those high-level adjustments.\footnote{Analogues of regulatory connections from high-level control structures have also been proposed in brain \cite{iyer2022avoiding, hawkins2016neurons, antic2018embedded}, furthering the relevance of this principle to AI.} Over time, these learned higher-order control flows or structures can themselves become encapsulated as new core processes, enabling the development of a multi-level, hierarchical organization within the learned model. 

\textit{\textbf{Contrast:}} In addition the points already discussed in (1), this also contrasts learning a connection solely between raw inputs and raw outputs in the black-box interpretation of neural networks.

\textbf{3) Growth and Local Variation \& Selection to Preserve Existing Knowledge:} Learning in biological systems occurs through mechanisms of local variation and selection, generating variation where needed to facilitate adaptive growth. AI architectures should implement similar local adaptation strategies, enabling dynamic complexity in models without disrupting established knowledge. As discussed in Section \ref{sec:edb_process_varsel}, this approach is crucial for creating new capacities—essentially a localized variation pool—where necessary, allowing for the acquisition of new knowledge without interfering with existing structures. Moreover, structural properties that facilitate adaptation, such as encapsulation and weak linkage, should not be imposed in a top-down manner but rather be created bottom-up through developmental principles from the lowest levels of organization, ensuring their expression at every level and enabling scalable, adaptive structural organization.

\textit{\textbf{Contrast:}} This contrasts not only with neural networks' lack of mechanisms for generating new structural components or sources of variation when needed for learning (see Sections \ref{sec:contemporary_ai} and \ref{sec:edb_process_varsel}), but also with the top-down design approach in ML literature which seeks to impose structure and properties such as weak linkage across modules of monolithic NNs from above \cite{su2024focuslearn, goyal2020object, pateria2021hierarchical}.

\subsection{Implications for AI research and early perspectives in the literature}
\label{sec:ai_future_principles}

The principles outlined in points (1) and (2), which can be regarded as key properties of structural organization, have been echoed in AI research and are widely recognized as desirable traits for AI systems. However, surprisingly little practical advancement has been made, given how intuitive these properties actually are. Many AI architectures impose rigid, predefined substructures that lack dynamism, rather than generating structured organization themselves \cite{su2024focuslearn, goyal2020object, pateria2021hierarchical}. Meanwhile, approaches that attempt to circumvent this constraint often rely on unstable foundations that restrict their expressive capacity due to inherent limitations in neural network construction. For instance, dynamic task-specific substructures in continual learning are fundamentally constrained by the task-dependency issues of NNs (see Sec. \ref{sec:ai_destructive_adaptation}) \cite{andreas2017modular, devin2017learning, sahni2017learning, goyal2019reinforcement, yang2020multi, pateria2021hierarchical, iyer2022avoiding}. Moreover, no existing approach achieves this self-organizing process at all levels of abstraction in a fractal-like manner.


The critical insight here lies in point (3): the integration of \textit{developmental principles}. These principles govern the structure of biological organisms and recursively-increasing pace of evolution, with reason to believe that they also underlie natural intelligence (Sec. \ref{sec:edb_process_varsel}). While structural organization and adaptive reuse (1 and 2) are desirable as \textit{sub-means} toward system-level continual learning and comprehensibility, they are not the \textit{fundamental means} themselves. The fundamental means to adaptability instead are rooted in the principles of low-level developmental mechanisms (3). AI research must acknowledge this and shift its focus accordingly. Without developmental principles—such as learning through local variation-selection instead of a universal gradient signal, or weak linkage at lower levels of organization instead of densely interconnected, fine-tuned units—operating at the foundational level to generate structure rather than imposing it from above, it becomes impossible to achieve the desired structural properties \textit{adaptively} and \textit{across all levels of organization}, the latter being particularly necessary for unbounded recursive improvability. Early works in this direction \cite{Erden2024Modelleyen, erden2025agential_extendedabs, erden2025agential, erden2025mnr} have demonstrated promising results by leveraging developmental principles, illustrating how these concepts can be integrated into AI systems to address key challenges such as the destruction of existing knowledge and incomprehensibility of learned representations. Advancing research in this direction and properly viewing this design paradigm as a whole—rather than as properties that can be realized in isolation—is essential.\footnote{The argument we present should not be mistaken for the view that AI systems must evolve a "genome" and express it as a final system, analogous to the genotype-phenotype transition in biological evolution. Several approaches—some dating back decades—have explored this concept, either by indirectly encoding the final structure through an adaptive genome \cite{meli2021study, gauci2010indirect}, or by evolving a developmental program or controller that constructs the final system \cite{gruau1996comparison, gruau1994automatic, de2015evolving, wilson2022evolving, najarro2023towards, zhang2024evolved}. While these methods integrate aspects of evolutionary developmental biology into AI, they differ from our perspective: they treat the encoding itself as the locus of adaptation and adapt it mostly throughout an external evolutionary process (although recent works examine alternatives, e.g. \cite{najarro2023towards}), whereas we discuss the promise of directly incorporating adaptation principles into the system-level learning process. These genome-based approaches offer significant advantages for applications that rely on or are computationally compatible with evolutionary computation, whereas the approach we advocate comes with the benefits of being more computationally practical (as it does not need an expensive evolution-unfolding-evaluation loop) and enabling a more direct and controllable integration of developmental principles into learning. While evolutionary developmental biology is closely linked to genotype-phenotype mapping, its core principles of adaptation \textit{do not} inherently require a system to be structured around this dichotomy.}

Complementarily, the complexity of core control processes in biological systems indicates that the appropriate design level for adaptive systems is \textit{not} high-level integration, but rather \textit{more proficient low-level organizational units}. This stands in contrast to the majority of current ML research, which focuses primarily on high-level mechanisms operating on networks of artificial neurons \cite{wan2024towards, colelough2025neuro} or building block alternatives with qualitatively similar capabilities \cite{bal2024rethinking}. There is limited exploration into fundamentally redesigning the capabilities of these core building blocks. It is unlikely that the design principles outlined above can be applied across all levels of organization using fundamental entities as simple as artificial neurons, underscoring the need for research in redesigning our core computational units to align with the developmental capabilities as discussed previously. Note that more capable computational building blocks will likely eliminate the necessity for complex integration mechanisms layered atop the fundamental learning algorithm, as their objectives will now be achieved by the low-level units—this should hence be viewed as a shift in the relevant design level (toward lower levels) rather than the introduction of new design goals. For example, see \cite{erden2025agential_extendedabs, erden2025agential}, where continual learning is embedded as an intrinsic property of a sophisticated low-level computational unit, eliminating the need for additional high-level mechanisms to prevent destruction of old knowledge, as commonly required in NN-based approaches

A possible concern regarding developmental processes may be computational viability. From the perspective of model size, we can be sure of feasibility—neural networks are already overparameterized for the tasks they learn \cite{du2018power}. In contrast, with proper minimal-growth methods, it is possible to construct a model that does not exceed the model complexity required by the task, aside from possible transients of local variation (Sec. \ref{sec:edb_process_varsel}). Concerning runtime, while individual learning steps would indeed exceed those of NNs, the capacity for continual learning (entailing online learning) removes the necessity for repetitive passes through the dataset (see \cite{erden2025agential, erden2025mnr} for a demonstration), which are the main culprits for the prolonged training sessions of traditional NNs. Notably, this stepwise-slower yet online learning regime is what is seen in animals, with even humans displaying nothing like the instantaneous gradient updates in NNs. Hence this approach is computationally feasible from both a size and runtime standpoint. This is not to say that feasibility will be immediately evident (e.g. \cite{erden2025agential_extendedabs, erden2025mnr} both run into some size issues) but it means that there are no insurmountable limitations beyond those already imposed by current methods.

\subsection{Integrating symbolic processes with multi-level structured representations}
\label{sec:ai_future_highlevel}

One way to interpret the insights from EDB is as an explanation for the generation of the organism’s internal multi-level organizational structure (not to be confused with multi-level selection theories \cite{damuth1988alternative, Okasha2005multilevel, gardner2015genetical, czegel2019multilevel}). When applied to AI, this concept of a multi-level structure—characterized by hierarchy, modularity, and reuse of common substructures—can be seen as a method for generating \textit{abstract} representations at multiple scales. These representations might include high-level percepts (e.g. features and objects), equivalence sets (e.g. percepts with shared outcome or alternative outcomes of the same percept), or behavioral patterns (e.g. subpolicies \cite{bakker2004hierarchical, li2019sub}). In addition to their potential for introducing comprehensibility to learned models (conditioned on appropriate analysis and interpretation of their internal composition, just like any hierarchically-organized software), such abstract representations can enable a more separable and precise depiction of the knowledge needed for high-level cognitive processes are notoriously difficult to integrate into modern learning systems that function as black boxes between raw observations and low-level actions/outcomes. Among these areas, the integration of learning with deliberative behavior (planning and decision-making with explicit constraints) and active information seeking stands out as particularly crucial.

As detailed in Section \ref{sec:ai_incomprehensible}, there is no effective method for integrating goal-oriented deliberative behavior with learning capabilities of contemporary ML. This integration becomes feasible, however, through structured representations generated by learning—abstract states, goals, and so on—allowing the identification of high-level conditions or outcomes, and linking them to abstract subgoals or subpolicies. A demonstration of the preliminary feasibility of an approach in this spirit can be found in \cite{erden2025agential_extendedabs, erden2025agential}, which integrates a basic learning system based on these principles with planning, as well as demonstrating a simple behavior encapsulation process.

Active information seeking naturally requires goal-oriented deliberative behavior, as the agent must deliberately act to achieve its information-seeking goals. However, it even further benefits by a multi-level structured representation as it also allows for an explicit representation of uncertainty and knowledge like significance or succession-probability regarding abstract entities like percepts or behavioral patterns, facilitating directed exploration and model updates based on this quantification. Active information seeking is already an active area of research in NN-based methods like reinforcement learning \cite{zhao2024active, wu2023uncertainty, mazumder2022knowledge}, yet its impact remains limited due to the challenges inherent in neural networks, as discussed throughout this article. The general approach behind the cited methods, however, could be applied to a structured multi-level representation, offering precision, intuitiveness, and scalability.

It's important to note that both deliberative behavior and active information seeking are already well-established fields. Deliberative behavior and planning are mature fields of research that have been active since the earliest days of computing \cite{ghallab2016automated}, and information seeking is grounded in probability theory and statistical inference \cite{parr2022active, zhao2024active, wu2023uncertainty, mazumder2022knowledge} dating back centuries. Numerous approaches in both fields can be seamlessly integrated into learning systems once a proper structured representation is in place. These integrations would transform (learning) AI systems, moving them beyond the limitation of learning solely from data or experience. Instead, these systems could leverage general algorithms for deliberative behavior, reasoning, information seeking, etc. as well as specific designer knowledge for their domain, achieving the integration sought by fields like neurosymbolic AI \cite{wan2024towards,colelough2025neuro} in an organic manner that applies across all organizational levels of the system.

\section{Conclusion}

The current design paradigm for AI, while widely adopted and successful in solving problems previously deemed unsolvable, has inherent limitations in essential qualitative capabilities expected from effective learning systems. These limitations impede AI’s potential, most notably their inability to learn new knowledge without losing previously acquired information. Additionally, the unstructured and overparameterized internal representations generated by current machine learning approaches prevent comprehensibility and integration with non-learning processes like deliberation or mechanisms of active information acquisition. This current state of AI parallels Modern Synthesis, the dominant perspective of evolution in 20th century. Recent advancements in evolutionary theory, particularly in evolutionary developmental biology, have addressed the shortcomings of the Modern Synthesis, offering a more comprehensive understanding of life’s evolution on Earth and better explanations for evolutionary patterns. Translating the principles of EDB into AI design could initiate a new paradigm that effectively overcomes the limitations of existing systems as a whole.





\newpage

.

\newpage

\appendix

\section{Appendix: Is intelligence evolutionary?}

In Section \ref{sec:edb_process_varsel}, we discussed exploratory processes—mechanisms operating within the organism that rely on somatic and local variation followed by selection. These processes exemplify the widely-seen phenomenon of Darwinian evolutionary mechanisms at a sub-population scale \cite{gerhart2007theory, marc2005plausibility, west2003developmental}, enabling the organism to generate adaptive responses to unpredictable circumstances. Such responses cannot be pre-encoded efficiently within the organism through genetic evolution, given the longer timescales required for population-level evolutionary change.

As discussed above, exploratory processes are widely observed in the formation and adaptation of the nervous system. Synaptic and axonal overproduction, followed by selective retention, occurs extensively during development and childhood \cite{hiesinger2021self, nihDevelopingBrain, huttenlocher2013synaptogenesis}. These processes persist into adulthood under conditions demanding high plasticity, albeit on a much smaller scale and in more localized regions \cite{gu2013neurogenesis, gonccalves2016vivo, mowery2023adult}. The widespread occurrence of these Darwinian mechanisms in neural development suggests they play a critical role in the emergence of intelligence.

Yet, this is not the whole story of evolutionary principles and their role in brain function. Since the mid-20\textsuperscript{th} century, numerous theories have been proposed to explain not only the local adaptive behavior of neurons but also higher-order brain functions through evolutionary principles. Among the most well-known is the theory of neural group selection \cite{edelman-neuraldarwinism-1987, edelman1993neural}, which posits a two-stage variation-and-selection process: first across neuronal connections and then across neuronal groups. Another line of research theorizes that neuronal activity patterns can implement evolutionary replicator dynamics \cite{fernando2010neuronal, de2015neuronal, fedor2017cognitive}, while other theories propose that synapses themselves act as self-interested agents \cite{seung2003learning} or as units of selection akin to genes \cite{adams1998hebb}, alongside additional formulations that explore alternative selectionist frameworks \cite{changeux1973theory, loewenstein2010synaptic, calvin1987brain, calvin1998cerebral, fernando2012selectionist}. While these ideas remain relatively speculative due to both the difficulty of experimental validation and the broad, sometimes imprecise, nature of their formulations, they nonetheless stand as plausible candidate theories for explaining brain function. This plausibility is reinforced by the well-documented presence of local evolutionary mechanisms at the neuronal level (such as synaptic and axonal overproduction followed by pruning, as described above) and the selectionist characteristics observed in certain high-level cognitive functions—such as selective attention \cite{johnston1986selective}.

Evolutionary mechanisms undeniably play a fundamental role in the emergence of intelligence in biological systems and are also implicitly embedded in the most effective artificial learning methods employed today. While we do not yet fully understand how these processes integrate into higher-level brain function or the extent to which they determine complex cognitive processes, we do know that somatic evolutionary mechanisms play an indispensable role in at least some aspects of cognition—particularly in learning. This necessity arises from the fundamental uncertainty regarding the future that any intelligent agent—whether an organism or an AI system—inevitably faces. For any given observation or circumstance encountered at a particular moment, the agent can generate multiple plausible solutions or internally consistent models/explanations, as its representational capacity inherently exceeds the requirements of any single observation (given that its cognitive model must accommodate more than just the present instance). Among these alternative solutions, many may be \textit{neutral} with respect to each other \cite{wilke2001adaptive, tenaillon2020impact} meaning they perform similarly well in the present context but differ in their suitability for future scenarios. Since an agent cannot a priori determine which of these neutral alternatives will prove most beneficial in the long run, the natural resolution to this problem is a mechanism of variation and selection. Multiple solutions of comparable immediate utility are generated and explored in parallel \cite{wagner2011origins}, and over time, those that prove most effective in future contexts are retained, while the rest are discarded.

\end{document}